\documentclass{article}

\usepackage[applemac]{ inputenc}%Usar este pacote de acento no mac

\usepackage{mathrsfs}
\usepackage{amsmath}
\usepackage{amssymb}
\usepackage{multicol}
\usepackage{float}
\usepackage{makeidx}
\usepackage{layout}
\usepackage{array}
\usepackage{boxedminipage}
\usepackage{hyperref}
\usepackage{latexsym}
\usepackage{indent first}

\usepackage{subcaption,graphicx}

\usepackage{color}

\usepackage[sectionbib,square]{natbib}%Pacote para definir novas formas de citacao

\usepackage[normalem]{ulem}%Pacote para sublinhar o texto de diversos modos

\title{Trapped solitary-wave interaction for Euler equations with low pressure region}
\author{M. V. Flamarion$^{1}$ and R. Ribeiro-Junior$^{2}$}
\date{}
\begin{document}

\maketitle

{\footnotesize
	\begin{center}
	$^{1}$ UFRPE/Rural Federal University of Pernambuco, UACSA/Unidade Acad{\^e}mica do Cabo de Santo Agostinho, BR 101 Sul, 5225, 54503-900, Ponte dos Carvalhos, Cabo de Santo Agostinho, Pernambuco, Brazil.

	$^2$ UFPR/Federal University of Paran\'a,  Departamento de Matem\'atica, Centro Polit\'ecnico, Jardim das Am\'ericas, Caixa Postal 19081, Curitiba, PR, 81531-980, Brazil. 
	
	\end{center}
	
	}

\begin{abstract}
Trapped solitary-wave interaction is studied under the full Euler equations in the presence of a variable pressure distribution along the free surface. The physical domain is flattened conformally onto a strip and the computations are performed in the canonical domain. Computer simulations display solitary waves that remain trapped in a low pressure region. In terms of confinement we observe that these waves are stable for small perturbations of either their amplitudes or the pressure forcing term. Furthermore multiple solitary waves  are considered within the low pressure region without escaping the low pressure region. We identify regimes in which multiple solitary waves remain trapped after several collisions. In particular we display a regime where three solitary waves are trapped and collide several times, before one escapes at a time. The remaining solitary waves stays trapped in the low pressure region. \\


\end{abstract}

\section{Introduction}
Waves generated by an external force frequently occur in nature or in laboratory experiments. These waves have been extensively studied through reduced models such as the Boussinesq and the forced Korteweg-de Vries (fKdV) (\cite{Wu, Paul, Binder, Marcelo-Paul-Andre}). Physically these waves can arise in the context of atmospheric flows encountering topographic obstacles or flow of water over rocks (\cite{Baines}). In addition, waves can also be generated when one considers a pressure distribution moving over a free surface, as for instance, ship waves and  ocean waves generated by storms (when a low pressure region moves on the surface of the ocean) (\cite{Johnson}). In a laboratory experiment performed by \cite{Pratt}, in which two obstacles were fixed and water  circulated through a flume with a variable-speed, it was observed a steady wave train between the two obstacles (see Pratt's figure $4$). Here we use the term ``trapped wave" to describe waves that remain trapped in a certain region, generally in between two topographic obstacles or in low pressure regions. When the trapped waves escape after sufficient time has elapsed we use the terminology temporally trapped waves.

From the fKdV point of view a great deal of attention has been paid on trapped waves over a bottom topography. For the interested reader, we mention a few recent references form which the bibliography may be useful. \cite{Grimshaw} showed that steady trapped waves can be found in the vicinity of a holelike forcing (topography) and studied numerically and theoretically how the shape of the hole affects the formation of trapped waves. Considering a bottom topography with two bumps \cite{Lee1} and \cite{Lee2} found steady waves and used them as initial data for the fKdV equation observing that those waves remained trapped between the two bumps up to a certain period of time. They also investigated the stability of those temporally trapped waves for small perturbations in their amplitudes and in the forcing term. 
In the same spirit \cite{Kim} studied the propagation of temporally trapped waves between two bumps submerged and showed that the trapped waves have to overcome a certain threshold of energy in order to go over the bumps.  They concluded that the energy barrier does not depend on the distance between the bumps, while it increases with the height of the bumps.

To the best of our knowledge there are no articles about time-dependent trapped waves solutions for the Euler equations. In particular, we here present the first study regarding waves trapped by a pressure distribution.  For steady trapped waves a study was done by \cite{Broeck}, under the assumption of a two-bump topography. They found  steady  wave trains in between the two obstacles and showed that its shape depends on the height and width of the bumps. They also showed that as the distance between the obstacles increases the flow approaches the generalised hydraulic fall past a single bump.

In this work  we consider the full Euler equations in the presence of a localised low-pressure distribution on the free surface as a proxy for a storm over a region. We found time-dependent trapped solitary waves which move back and forth in the low pressure region).  We also investigate the stability of these waves for small perturbations of their amplitudes and the profile of the pressure profile with respect to its amplitude and range. Our numerical experiments indicate that single solitary waves are stable, namely remaining trapped. An additional numerical result is the existence of multiple trapped solitary waves. We have found regimes in which several solitary waves remain trapped in the low pressure region after many collisions.

The paper is organized as follows. In section $2$ we present the mathematical formulation of the non-dimensional Euler equations. In section $3$  we rewrite the Euler equations in the canonical domain, which is a uniform strip where computations are more easily performed and present the numerical method to solve the equations. The results are presented in section $4$ and the conclusion in section $5$.
\section{The Euler equations in the presence of a moving pressure}
We consider a two-dimensional incompressible and irrotational flow of an inviscid fluid with constant density ($\rho$) in a channel of finite depth ($h_0$) in the presence of gravity ($g$) with a traveling pressure distribution with constant speed ($U_0$) on the free surface. We denote  the velocity potential by  $\tilde{\phi}(x,y,t)$  and the free surface profile by $\tilde{\zeta}(x,t)$. Choosing  $h_0$, $(gh_0)^{1/2}$, $(h_0/g)^{1/2}$ e $\rho g h_0$ as our characteristic scales for length, velocity, time and pressure, respectively, the non-dimensional governing equations are as follows:
\begin{align} \label{euler}
\begin{split}
& \Delta{\tilde{\phi}}= 0 \;\  \mbox{for} \;\ -1 < y <{\tilde{\zeta}}(x,t), \\
& {\tilde{\phi}}_{y} =0\;\ \mbox{at} \;\ y = -1, \\
& {\tilde{\zeta}}_{t}+{\tilde{\phi}}_{x}{\tilde{\zeta}}_{x}-{\tilde{\phi}}_{y}=0
\;\ \mbox{at} \;\ y = {\tilde{\zeta}}(x,t), \\
& {\tilde{\phi}}_{t}+\frac{1}{2}({\tilde{\phi}}_{x}^2+{\tilde{\phi}}_{y}^{2})+{\tilde{\zeta}}= - P(x+Ft) \;\ \mbox{at} \;\ y = {\tilde{\zeta}}(x,t),
\end{split}
\end{align}
where $F=U_0/(gh_0)^{1/2}$  is the Froude number. It is convenient  to consider the equations in the moving frame $x\rightarrow x+Ft$ and by writing
\begin{equation*}
\tilde{\zeta}(x-Ft,t)=\bar{\zeta}(x,t), \;\ \tilde{\phi}(x-Ft,y,t)=\bar{\phi}(x,y,t),
\end{equation*}
we obtain
\begin{align} \label{eu1}
\begin{split}
& \Delta{\bar{\phi}}= 0 \;\  \mbox{for} \;\ -1 < y <{\bar{\zeta}}(x,t), \\
& {\bar{\phi}}_{y} =0\;\ \mbox{at} \;\ y = -1, \\
& {\bar{\zeta}}_{t}+F{\bar{\phi}}_{x}{\bar{\zeta}}_{x}-{\bar{\phi}}_{y}=0
\;\ \mbox{at} \;\ y = {\bar{\zeta}}(x,t), \\
& {\bar{\phi}}_{t}+F{\bar{\phi}}_{x}+\frac{1}{2}({\bar{\phi}}_{x}^2+{\bar{\phi}}_{y}^{2}) +{\bar{\zeta}}= - P(x) \;\ \mbox{at} \;\ y = {\bar{\zeta}}(x,t).
\end{split}
\end{align}
We are interested in investigating the scenario in which the wave is trapped in a low pressure region. According to our adimensionalization $P=0$ means atmospheric pressure, while positive pressure means a high pressure region and negative pressure means a low pressure region. 

\section{ Conformal mapping and numerical methods}
In order to deal numerically with the system (\ref{eu1})  we proceed in the same fashion as \cite{Dyachenko}, which roughly speaking consists in constructing a time-dependent conformal mapping 
\begin{equation*}
z(\xi,\eta,t) = x(\xi,\eta,t)+iy(\xi,\eta,t),
\end{equation*}
which maps a strip of width $D$ onto the fluid domain, satisfying the boundary conditions 
\begin{equation*}
y(\xi,0,t)=\overline{\zeta}(x(\xi,0,t),t) \;\ \mbox{and} \;\ y(\xi,-D,t)=-1.
\end{equation*}
The canonical depth  $D$ is a function of $t$,  which is chosen so that the canonical and physical domains have the same length.
%As equacoes cinematica e Bernoulli $(\ref{eu1})_{3,4}$ no sistema de coordenadas canonicas podem ser escritas como: 

Let $\phi(\xi,\eta,t) =\bar{\phi}(x(\xi,\eta,t),y(\xi,\eta,t),t)$  and $\psi(\xi,\eta,t) =\bar{\psi}(x(\xi,\eta,t),y(\xi,\eta,t),t)$ where $\bar{\psi}$ is the harmonic conjugate of $\bar{\phi}$. Their traces along $\eta=0$ is denoted by  $\mathbf{\Phi}(\xi,t)$ and $\mathbf{\Psi}(\xi,t)$. By $\mathbf{X}(\xi,t)$, $\mathbf{Y}(\xi,t)$ we denote the horizontal and vertical free surface coordinates at $\eta=0$. Substituting these variables in $(\ref{eu1})_{3,4}$ yields when evaluated
\begin{align}\label{eulerconforme}
\begin{split}
& \mathbf{Y}_{t} =\mathbf{Y}_{\xi}\mathcal{C}\bigg[\frac{\mathbf{\Psi}_{\xi}}{J}\bigg] 
-\mathbf{X}_{\xi}\frac{\mathbf{\Psi}_{\xi}}{J}, \\
& \mathbf{\Phi}_{t} = - \mathbf{Y} - \frac{1}{2J}
(\mathbf{\Phi}_{\xi}^{2}-\mathbf{\Psi}_{\xi}^{2}) +\mathbf{\Phi}_{\xi}\mathcal{C}\bigg[\frac{\mathbf{\Psi}_{\xi}}{J}\bigg] 
- \frac{1}{J}F\mathbf{X}_{\xi}\mathbf{\Phi}_{\xi} - P(\mathbf{X}),
\end{split}
\end{align}
where $D = 1+ \big<\mathbf{Y}(\cdot,t)\big>$,  $\mathcal{C}=\mathcal{F}^{-1}_{k\ne 0}i\coth(kD)\mathcal{F}_{k\ne 0}$, $J=\mathbf{X}_{\xi}^2+\mathbf{Y}_{\xi}^2$ and
\begin{align}
\begin{split}
& \mathbf{X}_{\xi} = \frac{1}{D}-\mathcal{C}\big[\mathbf{Y}_{\xi}\big], \\
& \mathbf{\Phi}_{\xi} = -\mathcal{C}\big[\mathbf{\Psi}_{\xi}(\xi,t)\big]. \\
\end{split}
\end{align}
Fourier modes are given by $$\mathcal{F}_{k}[g(\xi)]=\hat{g}(k)=\frac{1}{2L}\int_{-L}^{L}g(\xi)e^{-ik\xi}\,d\xi,$$
$$\mathcal{F}^{-1}_{k}[\hat{g}(k)](\xi)=g(\xi)=\sum_{j=-\infty}^{\infty}\hat{g}(k)e^{ik\xi},$$
where $k=(\pi/L)j$, $j\in\mathbb{Z}$. More details about these computations can be seen in \cite{JFM17} and \cite{Marcelo-Paul-Andre}. Moreover, the adimensional mass  integral quantity in the canonical system is given by the formula

\begin{equation}
M(t) = \int_{-L}^{L}\mathbf{Y}(\xi,t)\mathbf{X}_{\xi}(\xi, t)\,d\xi.
\end{equation}\label{massa}

In the following sections we consider the computational domain $[-L,L]$, with an uniform grid with $N$ points and step $\Delta \xi = 2L/N$. All derivatives in $\xi$ are computed spectrally (\cite{Trefethen}).  Besides, the time evolution of the system (\ref{eulerconforme}) is calculated through the Runge-Kutta fourth order method with time step $\Delta t$. 
Unless mentioned otherwise, we use the following parameters: $ L = 750$, $N = 2^{13}$ and $\Delta t = 0.1$.
\section{Results}

In all simulations presented in this work we consider as initial data for system (\ref{eulerconforme}) unforced traveling wave solutions with velocity $c\approx 1.09$, namely computed in the absence of pressure ($P=0$) and Froude number $F=0$. The initial data is obtained in the same fashion as \cite {JFM17} by a Newton continuation method.

We verify our numerical method by first evolving the initial data considering $P=0$ and $F=-c$ in the system (\ref{eulerconforme}). The method is accurate in the sense that the solution $\zeta(x,t)$ is indeed a steady wave. Furthermore, denoting the initial wave profile by $\zeta(x,0)=\zeta_{0}(x)$ and by $$E_{A}=\max_{0\le t\le 35000}\max_{x}|\zeta_{0}(x)-\zeta(x,t)|,$$ the absolute error we get the relative error $$E_{R}=\frac{E_A}{\displaystyle\max_{x}|\zeta_{0}(x)|}=\mathcal{O}(10^{-11}).$$

%Trapped waves have been studied for several authors \cite{Kim, Lee1, Lee2, Broeck}. All these works have considered only steady trapped waves i.e. they computed steady waves, 
%and then investigated its stability. 

\begin{figure}[h!]
	\centering	
	\includegraphics[scale =0.9]{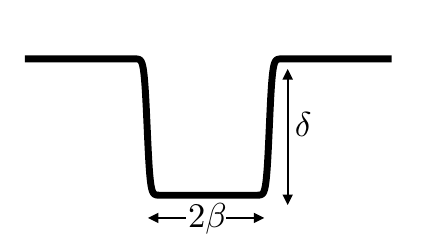}
	\caption{Profile of the pressure and its parameters.}
	\label{EsquemaPressao}
\end{figure}

Now, we study time-dependent trapped solitary waves in low pressure regions. For this purpose we assume that in system (\ref{eu1}) the pressure term $P(x)$ on the free surface is modelled by the function  $$P(x)=\delta(\tanh(x-\beta)-\tanh(x+\beta)),$$
with its profile is depicted in figure \ref{EsquemaPressao}. Unless mentioned otherwise, we assume $\delta=10^{-4}$, $\beta=20$ and $F=-c-0.01$.

\begin{figure}[h!]
	\centering
	\includegraphics[scale =0.8]{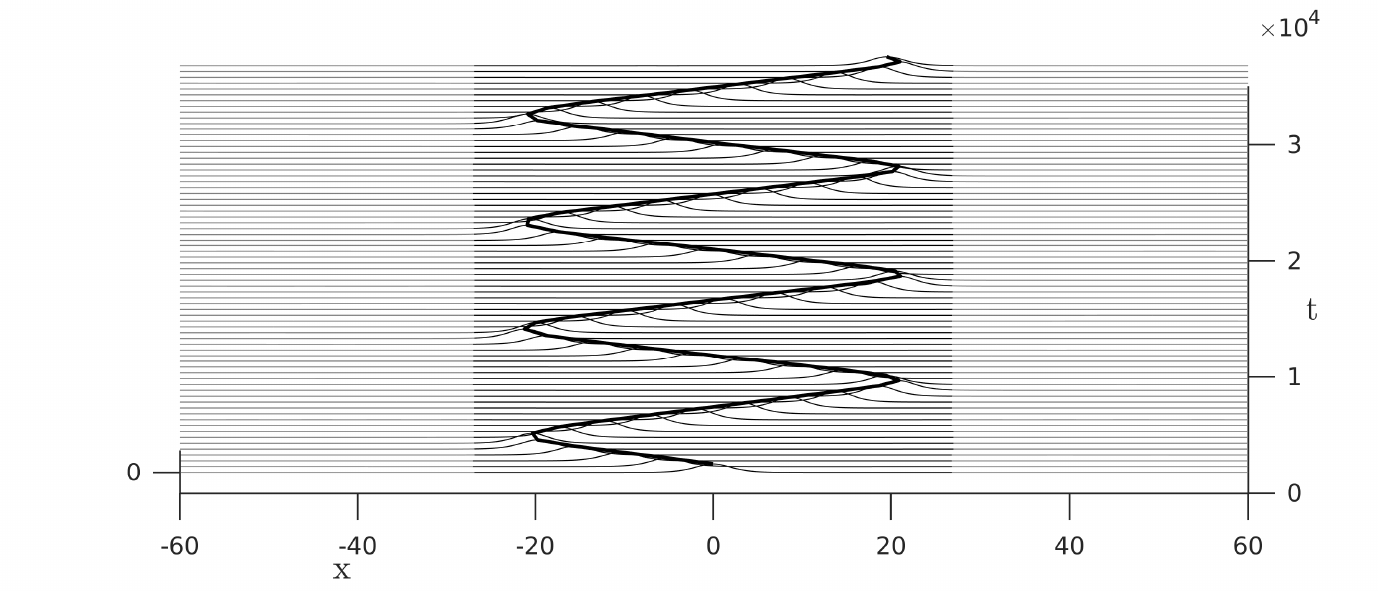}
	\includegraphics[scale =0.8]{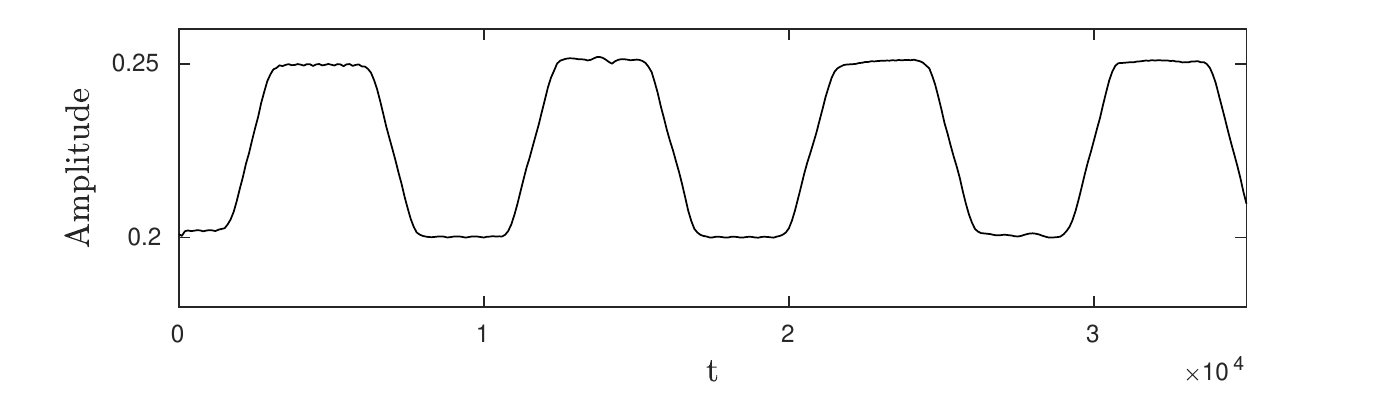}
	\caption{Top:  Trapped solitary wave in the low pressure region (darker area). The thicker black line shows the evolution of the crest of the wave. Bottom: The amplitude of the wave as a function of time.}
	\label{FigTrapped1}
\end{figure}

Figure  \ref{FigTrapped1} (top) shows the evolution of a trapped solitary wave in the low pressure region (darker region) as well as the dynamics of its crest (thicker black line). Initially the solitary wave crest is at $x=0$. As time elapses the wave travels downstream until it reaches  the high pressure region, then it reflects back  and moves upstream until it reaches the high pressure region again. This motion is repeated several times. Figure \ref{FigTrapped1} (bottom) displays how the amplitude of the trapped wave  varies throughout time. Notice that when the wave is moving downstream its amplitude is smaller than when it is moving upstream. Besides it is remarkable that the wave behaves effectively as a travelling wave. No radiation is observed during reflection and amplitude adjustments. The same will be observed during collisions as reported below. In this simulation, the  mass is conserved (\ref{massa}). More precisely, we have that
$$\frac{\displaystyle\max_{0\le t\le 35000}|M(t)-M(0)|}{|M(0)|}\le\mathcal{O}(10^{-10}).$$

%\subsection{Stability of trapped waves}

Some authors (\cite{Lee1, Kim, Lee2}) have studied the stability of trapped waves for the Forced-Korteweg-de-Vries equation by disturbing its amplitude and its forcing term  examining  whether the waves remain trapped or not. Along the same lines we disturb the amplitude of the wave showed on figure \ref{FigTrapped1}.

\begin{figure}[h!]
	\centering
	\includegraphics[scale =0.7]{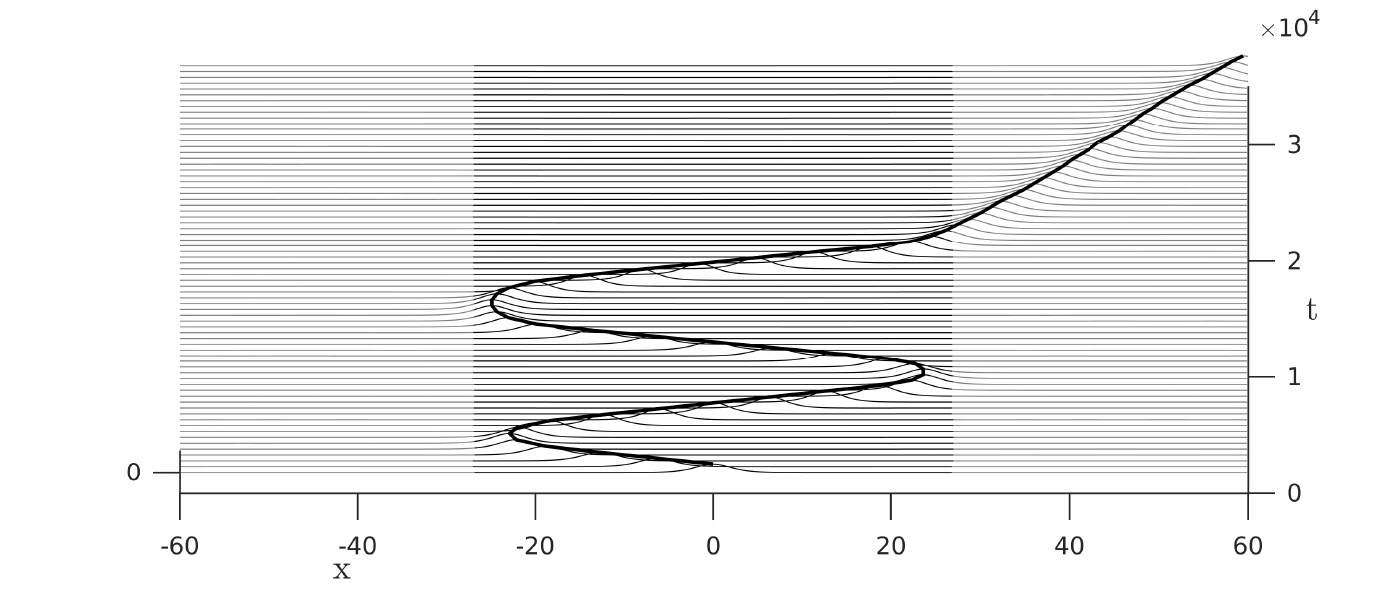}
	\includegraphics[scale=0.7]{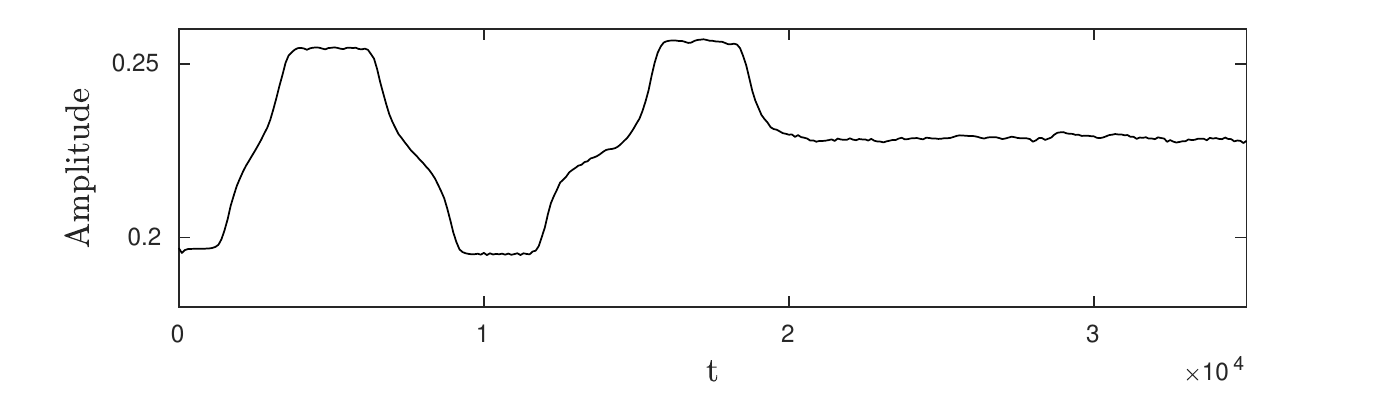}
	\caption{Top: The figure shows the evolution of the trapped solitary wave with $2\%$ reduction of its amplitude. The thicker line displays the propagation of its crest. Bottom: The amplitude of the wave as a function of time. }
	\label{2porcento}
\end{figure}

\begin{figure}[h!]
	\centering
	\includegraphics[scale =0.7]{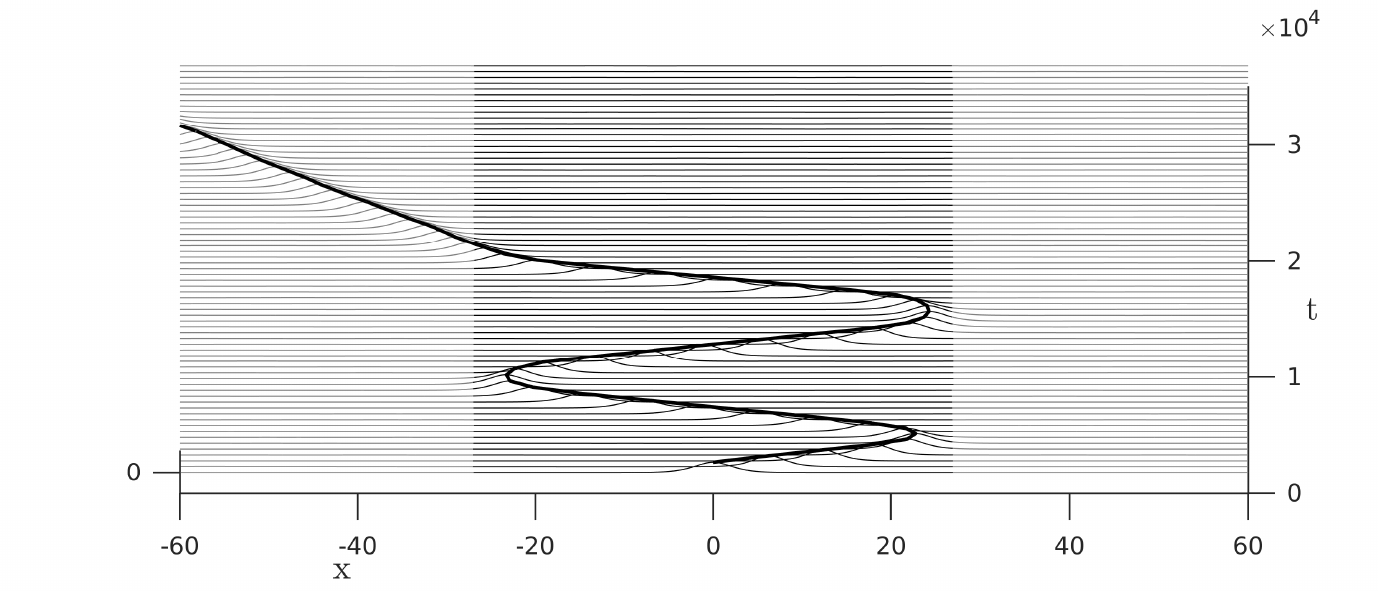}
	\includegraphics[scale=0.7]{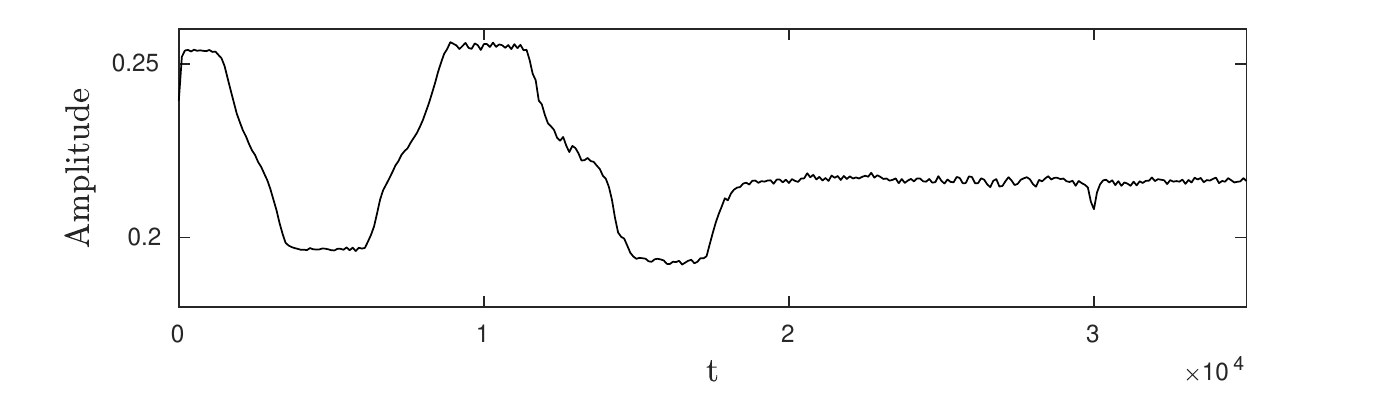}
	\caption{Top: The figure shows the evolution of the trapped solitary wave with $20\%$ accretion of its amplitude. The thicker line represents its crest propagation. Bottom: The amplitude of the wave as a function of time.  }
	\label{20porcento}
\end{figure}

Firstly, we consider the initial data under a perturbation with a $2\%$ reduction of its amplitude. The results are shown in figure \ref{2porcento}. The wave that was previously trapped, now escapes. In our numerical simulations we found that for reductions smaller than $2\%$ the wave never reverses its movement, i.e, it always moves downstream.
Secondly, we increase the amplitude of the trapped wave in $20\%$. This is depicted in figure \ref{20porcento}. As we can see
the perturbation is strong enough to make the wave move upstream at $t=0^{+}$ and leave the low pressure region at $t\approx 2\times 10^{4}$. We would like to point out that the wave remains trapped (at least until $t= 35\times 10^{4}$) for perturbations smaller  than $20\%$.
To the best of our knowledge there is no theory to explain the nontrivial behaviour of the temporally trapped solitary wave.

\begin{figure}[h!]
	\centering
	\includegraphics[scale =0.7]{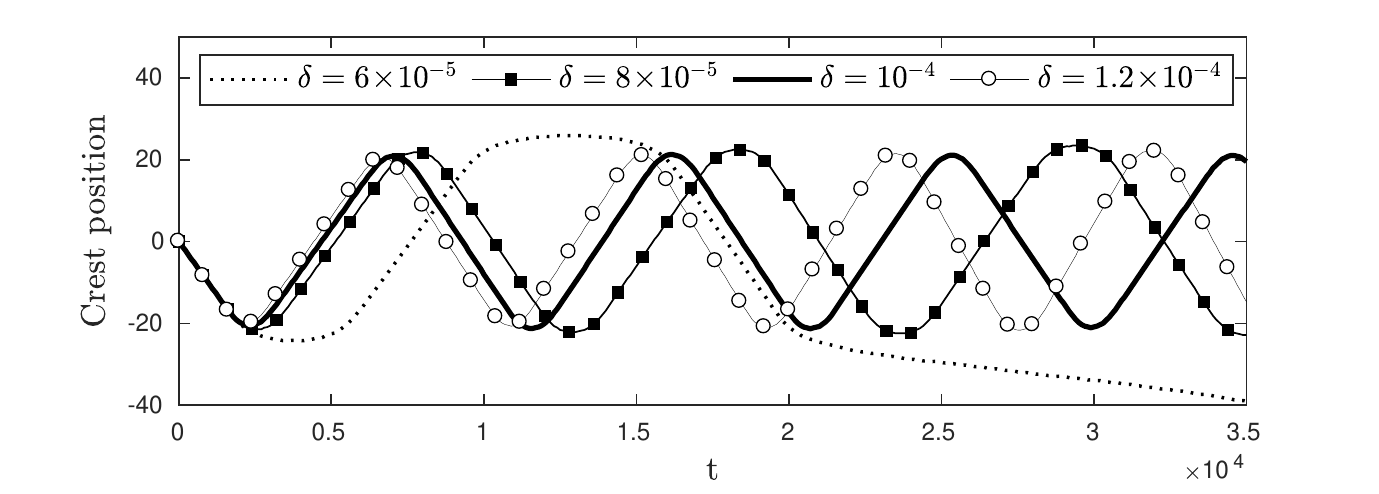}
	\includegraphics[scale =0.7]{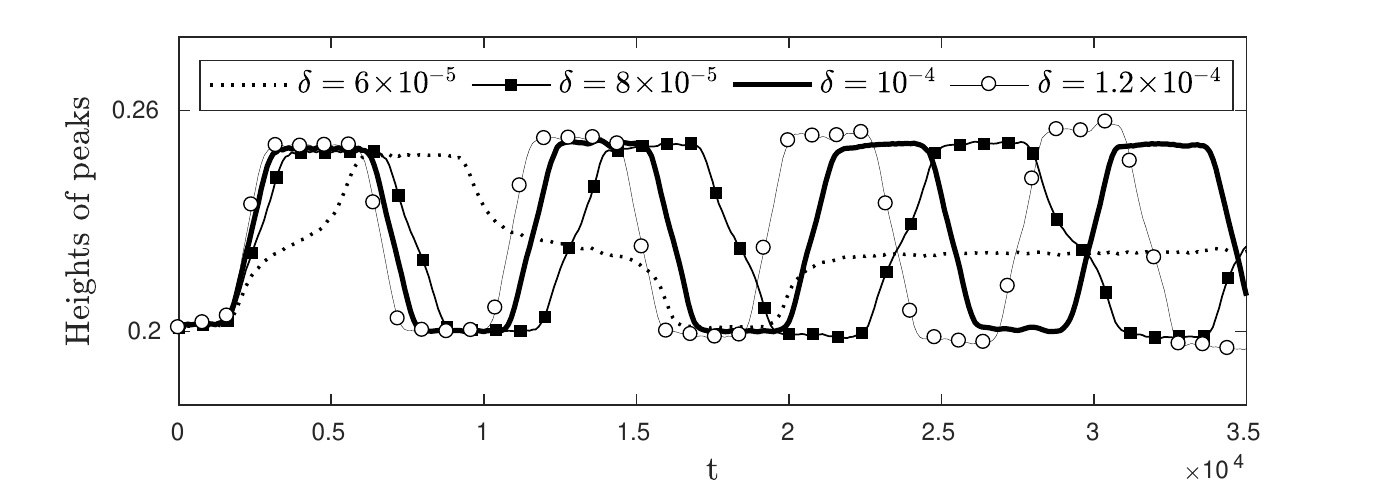}
	\caption{Top: Propagation of the crest for different values of $\delta$. Bottom: The heights of the highest peaks as function of time.}
	\label{crestposition}
\end{figure}

We also consider perturbations in the amplitude of the pressure $(\delta)$. In our simulations we observed that the trapped wave is robust for reductions below $20\%$.  We carried out accretions of $\delta$ up to $50\%$ and the wave remained confined in the low pressure region (at least until $t= 35\times 10^{4}$). For accretions above $50\%$ the generated waves due to the pressure may break  (see \cite{Grimshaw})  which drives this analysis to another subject that is not within the scope of our study. Figure \ref{crestposition} displays the propagation of the crest of some waves for different values of $\delta$ (top) and the heights their  peaks (bottom). On one hand as $\delta$ increases so does the back-and-forth motion of the trapped waves. On the other hand  the fluctuations of the peaks do not change much as time elapses. Note that the smallest value of delta is too weak to promote the oscillator pattern of the solitary wave.

Regarding of the parameter $\beta$, the initial wave remains trapped in the low pressure region up to $t=35\times 10^{4}$ for values of $\beta\ge 10$ (which is approximately half the efective wavelength of the initial wave).

\begin{figure}[h!]
	\centering
	\includegraphics[scale =0.9]{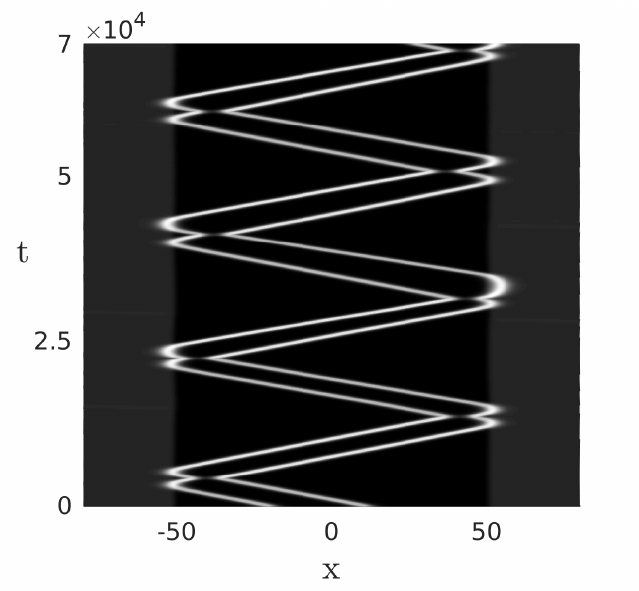}
	\includegraphics[scale =0.9]{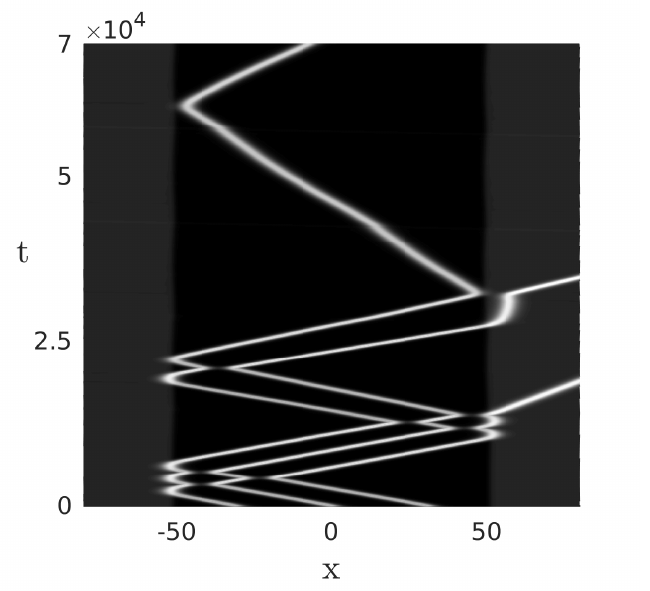}
	\caption{The figure shows the collision of solitary waves in the low pressure region. Left: Two solitary waves trapped in the low pressure area. Right: The propagation of three solitary waves.  As time elapses two of the waves escapes and only  one remains trapped. Parameters: $\beta=50$ and $\delta=10^{-4}$.}
	\label{Choques}
\end{figure}

Lastly we consider multiple solitary waves as initial data for (\ref{eulerconforme}) in a larger low pressure region. For this purpose we set the initial data as sums of translations of $\zeta_0$. Figure \ref{Choques} (left) displays two solitary waves propagating downstream at $t=0^+$. When the left solitary wave reaches the high pressure region it reflects back upstream and collides with the other solitary wave and both reverse  directions. Then this behaviour repeats on the right side of the high pressure region and we observe that after many collisions  both solitary waves remain trapped in the low pressure region. Figure \ref{Choques} (right) depicts three solitary waves. Initially all waves propagate downstream. Differently of the previous case after a few collisions one solitary wave escapes the low-pressure region. After another shorter series of collisions a secondary solitary wave escapes with a remaining trapped wave. To the best of our knowledge the  scenarios here presented have never been contemplated in the literature, not even for the fKdV equation.

\section{Conclusion}

In this paper we have studied time-dependent trapped solitary waves solutions for the full Euler equations. Through an iterative method we computed we computed solitary waves for the Euler equations and showed that for certain regimes these waves  move back and forth oscillating within  the low pressure region. Our numerical investigations demonstrated that the trapped wave solutions are stable for small perturbations of either their amplitudes or the forcing term. Furthermore we discovered regimes in which multiple solitary waves can temporally be trapped in a low pressure region. The numerical method applied in this work can be extended to the case in which there is a variable topography. This is a first step in studying time-dependent dynamics
for trapped waves under different types of forcing and different flow regimes. New solitary wave-interaction regimes were uncovered, with further investigations to be pursued in the future.

\section{Acknowledgements}

The authors are grateful to IMPA-National Institute of Pure and Applied Mathematics for the research support provided during the Summer Program of 2020 and to Prof. Andr{\' e} Nachbin (IMPA) for his constructive comments and suggestions which improved the manuscript. 
M.F.   is grateful to Federal University of Paran{\' a} for the visit to the Department of Mathematical Sciences. 
R.R.-Jr  is grateful to University of Bath for the extended visit to the Department of Mathematical Sciences.

\bibliographystyle{abbrv}

\end{document}